\documentclass[a4paper,11pt]{article}
\pdfoutput=1 

\usepackage{jcappub} 

\usepackage[T1]{fontenc} 

\usepackage{float}

\title{Angular resolution of stacked resistive plate chambers}

\author[a,1]{Deepak Samuel,\note{Corresponding author.}}
\author[a]{Pratibha B Onikeri,}
\author[a]{Lakshmi P Murgod}

\affiliation[a]{Central University of Karnataka,\\Kalaburagi, India}

\emailAdd{deepaksamuel@cuk.ac.in}
\emailAdd{pratibhaonikeri@gmail.com}
\emailAdd{lakshmipmurgod@gmail.com}

\abstract{We present here detailed derivations of mathematical expressions for the accuracy in the arrival direction of particles estimated using a set of stacked resistive plate chambers (RPCs). The expressions are validated against experimental results using data collected from the prototype detectors (without magnet) of the upcoming India-based Neutrino Observatory (INO). We also present a theoretical estimate of angular resolution of such a setup. In principle, these expressions can be used for any other detector with an architecture similar to that of RPCs.}

\begin{document}
\maketitle
\flushbottom

\section{Introduction}
\label{sec:intro}
Resistive plate chambers (RPCs) are particle detectors which have found applications in a wide range of high energy physics experiments. There are a variety of RPCs designed for specific purposes like triggering, timing and tracking. A review of its construction and working can be found in existing literature \cite{santonico1981development, fonte2002applications}. An arrangement of RPCs stacked together either horizontally or vertically can be used to track the trajectory of a particle as used in detectors like the Iron Calorimeter (ICAL) of the India-based Neutrino Observatory (INO) \cite{mondal2012india}.\\
An important parameter in tracking analysis is the accuracy of the arrival direction of particles and angular resolution of the detector setup, which are usually deduced from experimental data or monte-carlo analysis. In this work, we derive mathematical expressions for the arrival direction accuracy of such a stacked detector in terms of the known parameters like the strip width, the height of the stack and the separation between the RPCs. Such an expression can be used as an alternative to the conventional methods mentioned above and can aid in cross verification.
We also show the relationship between the accuracy in arrival direction and linear fitting parameters for cases where the track can be approximated by a straight line. This expression is validated against data acquired from a prototype stack of the INO-ICAL using the odd-even difference method developed on the basis of the chessboard method \cite{di2007measurement}. 

\section{Detector description and experimental setup}
\begin{figure}[htbp]
	\centering
		\includegraphics[scale=0.5]{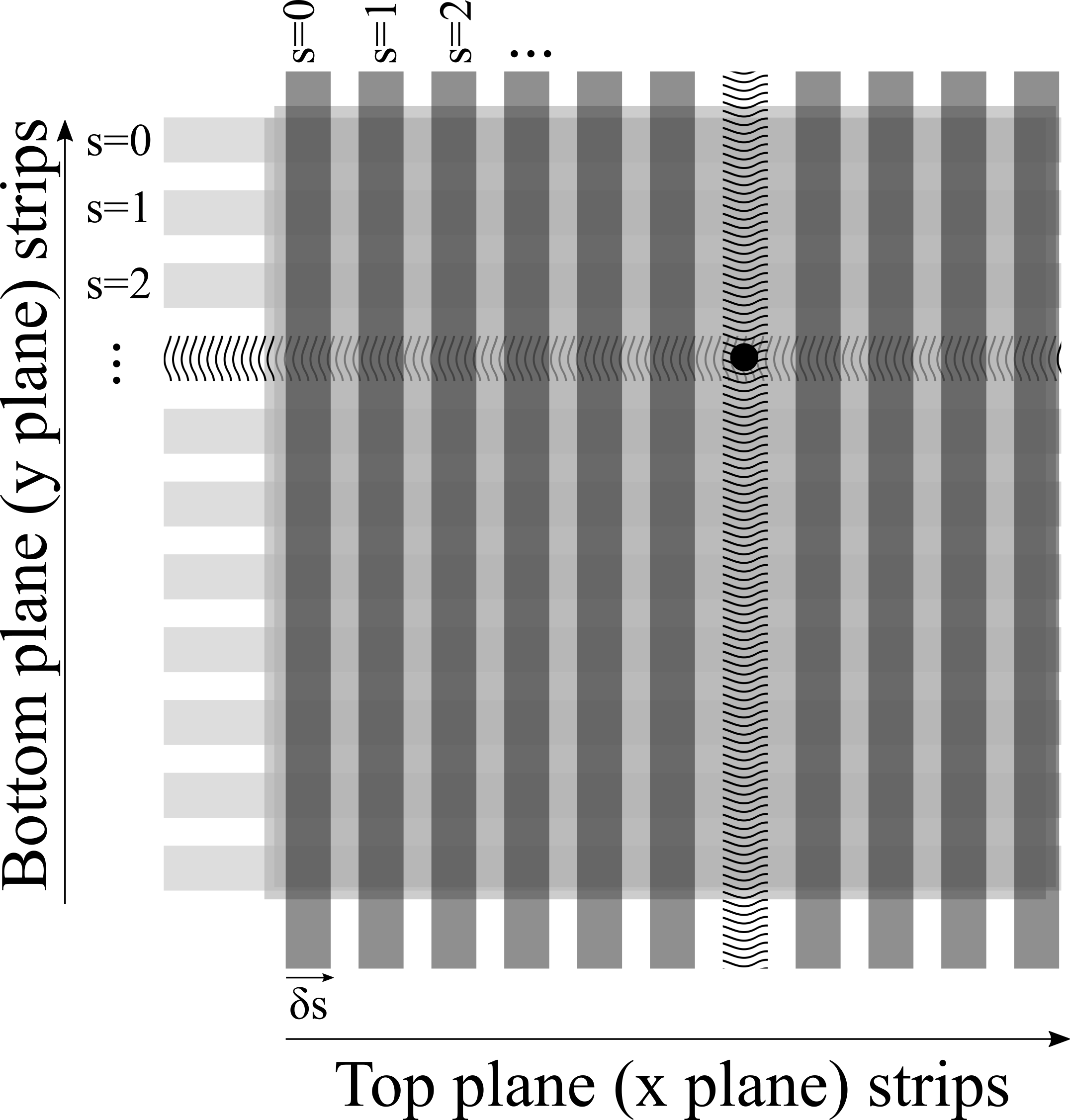}
	\caption{Schematic of a typical RPC  showing the two pickup planes, one on the top, here referred as x plane and another in the bottom, referred as y plane. The electrodes are sandwiched between these two planes. The black dot represents a particle crossing point and the strips on which the signals are induced due to the particle crossing are shown by wavy lines.}
	\label{fig:rpcbd1}
\end{figure}

\begin{figure}
				\centering
					\includegraphics[scale=0.5]{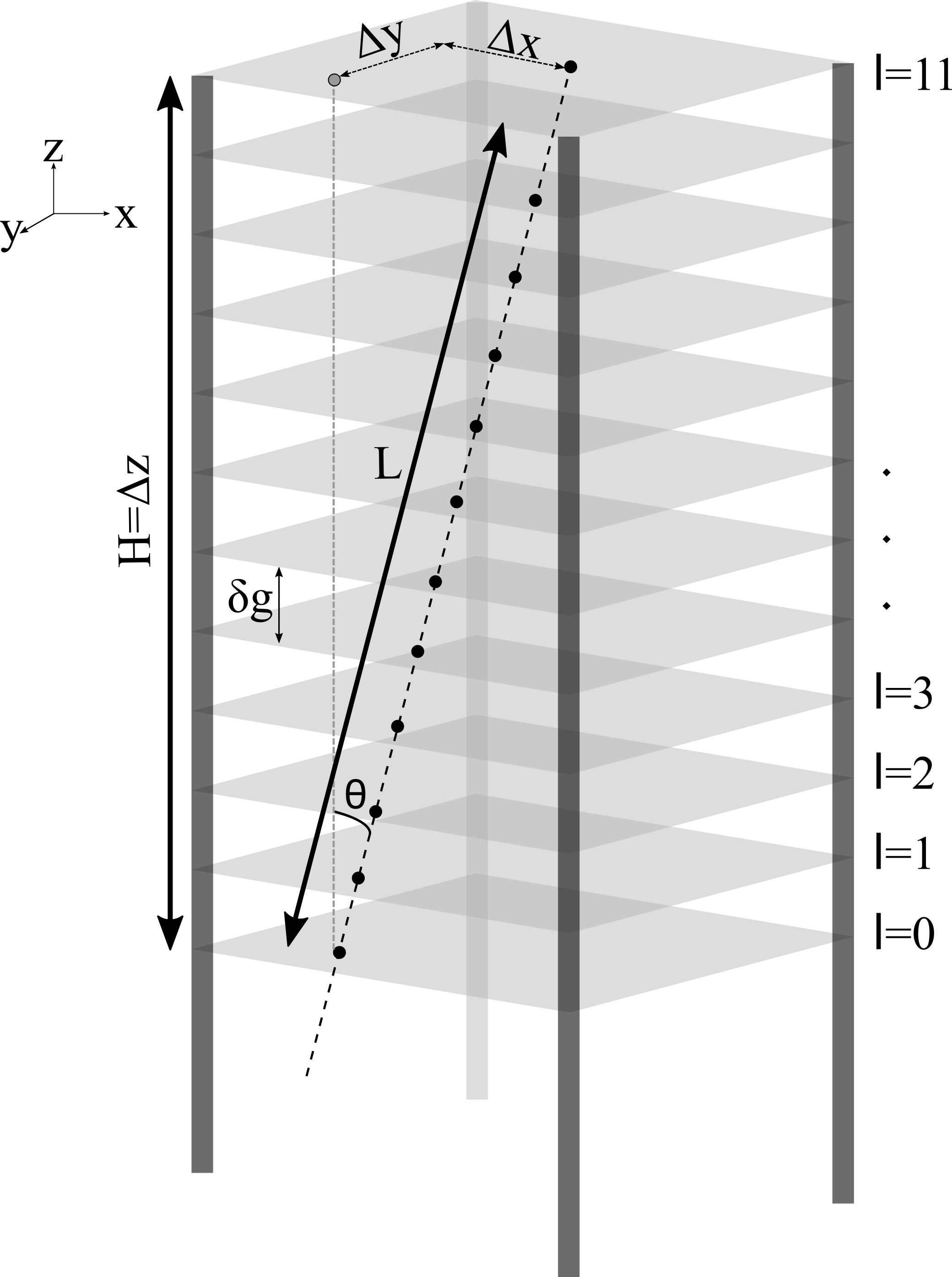}
				\caption{Schematic of a stack of RPCs showing the parameters used in the derivation of the expressions for accuracy in the arrival direction. The black dots are particle crossing points at each layer and the gray dot on the top layer is the projection of the crossing point on the bottom layer.}
				\label{fig:stack}
			\end{figure}

RPCs have two pickup planes: a top plane, here referred as the x plane and a bottom plane, referred as the y plane. These planes have strips of a specific width placed over them for signal collection. The strips on the x plane are orthogonal to the strips on the y plane. Two parallel electrodes, which encompass a gas mixture, are sandwiched between these planes (Cf. Figure \ref{fig:rpcbd1}). A particle crossing the detector will induce signals in one or more of the strips close to the crossing point both on the x and the y planes. These signals can be read out to identify the crossing point (x, y) of the particle in the plane of the detector. If multiple detectors are placed along the path of the particle, the trajectory of the particle can be obtained from the crossing points read out from these detectors, as shown in Figure \ref{fig:stack}. The expression derived in this work is for a typical configuration in which the spacing between the detectors (also called layers) is uniform and the strip width is constant.
            \\
As part of its detector development program, the INO collaboration has setup prototype RPC detector stacks at various institutions. The motivation for this program is to study and set the performance benchmarks for RPCs that will be used in the final detector. ICAL being primarily designed as a neutrino detector requires the RPCs to have high efficiency ($> 95\%$) and a timing resolution of better than 1 ns. These goals have already been reached in the 1 m x 1 m prototypes \cite{satya_thesis}.

This study was carried out using the data collected from the prototype stack (without magnet) at the Inter Institutional Centre for High Energy Physics (IICHEP) in Madurai, a city in the southern part of India.  This prototype detector stack has 8 layers of 1 m x 1 m glass RPCs. Each RPC is made of 2 glass plates (the electrodes) of thickness 3 mm separated by a gas gap of about 2 mm. A conductive coating is applied on the glass plates and a high voltage of about 10 kV is maintained across the electrodes. The RPCs are operated in the avalanche mode with a gas mixture of Freon, Isobutane and SF$_6$ in the ratio of 95.5 : 4.2 : 0.3. The pickup panels are placed in orthogonal directions over the glass electrodes which enables the readout out of the 2-d hit position of the particles. Each pickup panel has 32 strips with a pitch of 30 mm. These strips are connected to analog front-end electronics which amplify the signal and forward it to the digital front-end electronics the output of which is then read out and stored by the data acquisition system. The RPCs are separated by a gap of about 16.8 cm. The data acquisition system reads out the timing information, strip hit information from all the RPCs on receipt of a trigger. The trigger for this study was generated by a coincidence of top and bottom layers (i.e. l=7 and l=0 only). The trigger rate for this coincidence logic is about 10--15 Hz. The data acquisition also periodically monitors the noise rate of the RPCs. The prototype detector at IICHEP is similar in all aspects to the stack at the Tata Institute of Fundamental Research (TIFR), Mumbai. The only exception is that the IICHEP stack has 8 layers whereas the TIFR stack has 12 layers. A detailed description of the detector setup and data acquisition system at TIFR can be found in \cite{bhuyan2012vme}  and  \cite{behere2013electronics}.  There are still ongoing developmental activities in various fronts especially in the design of 2 m x 2 m RPCs, new gas mixtures and fast electronics.

\begin{table}[htbp]

	\centering
		\begin{tabular}{| l | l |}\hline
			Parameter & Description\\\hline\hline
			$\delta s$& strip pitch (strip width + gap)\\
			$N_s$& number of strips\\
			$\delta g$& gap between the detectors\\
			$N_l$& number of layers in the stack\\
			l& layer index (0 to $N_l$-1)\\
			s& strip index (0 to $N_s$-1)\\
			H& height of detector volume = $\delta g\cdot (N_l-1)$\\
			L& path length inside detector volume\\
			$\theta$&zenith angle\\\hline
		\end{tabular}
		\caption{The list of parameters used in the derivation of the accuracy of arrival direction}
		\label{param_table}
\end{table}

\section{Expression for accuracy of arrival direction}

The prototype stacks are typically used for testing and studying the detector and electronics characteristics. However, there were few physics studies like the measurement of zenith angle distribution and velocity of cosmic muons that have been undertaken in the past \cite{majumder2012velocity}. The angular resolution and the accuracy of the arrival direction of the final ICAL detector will be important parameters in many of its physics studies. In this work, we particularly focus on deriving a mathematical expression for accuracy of one component of arrival direction namely the zenith angle and compare it with the data from the IICHEP stack. We consider only the statistical part and ignore systematic errors like detector misalignment and also the effect of multiple scattering by the particles. Such effects require a monte-carlo study and is beyond the scope of the paper. The ICAL detector has Iron plates in between the layers and also a magnetic field which makes it different from the prototype setups. Therefore, these expressions cannot be used directly for estimating the accuracy of the arrival direction for the final detector. However, this study can serve as a guiding line and also provide an estimate of the limits of the accuracy in the arrival direction.    

The symbols used in the ensuing discussion are listed in Table \ref{param_table}. The parameter $l$ is an integer between 0 and N-1 that identifies a particular RPC. $s$ is an integer between 0 and $N_s-1$ that identifies a particular strip in a RPC. 

The zenith angle $\theta$, as shown in Figure \ref{fig:stack}, is the angle  made by the particle track with the normal to the detector plane, i.e.,:
\begin{equation}
cos (\theta) = \frac{H}{L} 
\label{costheta}
\end{equation}
where L is the path length inside the detector volume and H is the height of the detector volume.\\
The accuracy is thus reflected in the uncertainty in $cos(\theta)$. The variance of a function x(u,v,...) is given by the error propagation equation \cite{ku1966notes}:
\begin{equation}
\sigma^2_x = \sigma^2_u{(\frac{\partial x}  {\partial u})}^2 +  \sigma^2_v{(\frac{\partial x}  {\partial v})}^2+... 
\label{errprop}
\end{equation}
Therefore, the relative variance in $cos (\theta)$, which is a function of H and L, is given by:
\begin{equation}
\frac{\sigma^2_{cos(\theta)}} {cos^2(\theta)}= \frac{\sigma^2_H} {H^2} + \frac {\sigma^2_L} {L^2} \approx  \frac {\sigma^2_L} {L^2} 
\end{equation} 
The path length varies between different tracks. However, since the height of the detector volume is fixed and usually known to a good accuracy, the uncertainty in H can be ignored as a first approximation.\\
The path length L is given by:
\begin{equation}
L = \sqrt{\Delta x^2 + \Delta y^2 +\Delta z^2}
\label{L}
\end{equation}
$\Delta x$ ($\Delta y$) is the displacement along the x (y) direction as the particle traverses a length of  $\Delta z$ along the z direction. Here, $\Delta z$ is equal to the height of the detector volume H.\\
The variance $\sigma^2_L$ of the path length L is obtained by using the error propagation formula given in Equation \ref{errprop}: 

\begin{equation}
\sigma^2_L = \sigma^2_{\Delta x}  {(\frac{\partial L} {\partial \Delta x})}^2 + \sigma^2_{\Delta y} {(\frac{\partial L}{\partial \Delta y})}^2 + \sigma^2_{\Delta z}  {(\frac{\partial L}{\partial \Delta z})}^2
\end{equation}

The partial derivatives of L can be found using Equation \ref{L} and we get:
\begin{equation}
\sigma^2_L = {\frac{{\sigma^2_{\Delta x}  {\Delta x}^2 + \sigma^2_{\Delta y} {\Delta y}^2 + \sigma^2_{\Delta z}  { \Delta z}^2}}{{\Delta x}^2 +{\Delta y}^2 +{\Delta z}^2}}
\label{lerr}
\end{equation}
$\sigma^2_{\Delta z}$, the variance in the height of the detector volume (i.e., $\sigma^2_{H}$), is assumed to be negligible. Therefore, if the variance in $\Delta x$ is comparable to the variance in $\Delta y$, Equation \ref{lerr} can be written as:

\begin{equation}
\sigma^2_L = \left( \sigma^2_{\Delta x}\right) \frac{\Delta x ^2 + \Delta y^2 + \Delta z^2 - \Delta z^2}    {\Delta x^2 +\Delta y^2 +\Delta z^2}
\label{lerr2}
\end{equation}

and therefore,
\begin{equation}
\frac{\sigma^2_L}{L^2} = \left( \frac{\sigma^2_{\Delta x}}{L^2}\right) \left(1 - \frac{H^2}{L^2} \right) = \left( \frac{\sigma^2_{\Delta x}}{L^2}\right) sin^2(\theta)
\end{equation}
Rewriting L in terms of the known parameters $cos(\theta)$ and H and further noting that H can be written as $(N_l-1)\cdot \delta g$, we get,
\begin{equation}
\frac{\sigma^2_{cos(\theta)}} {cos^2(\theta)} = \left( \frac {\sigma^2_{\Delta x}}{(N_l-1)^2 \delta g^2}\right)\sin^2(\theta)cos^2(\theta) 
\label{theory} 
\end{equation}
All the factors except $\sigma^2_{\Delta x}$ in Equation \ref{theory} are known. The following section describes a method to estimate this quantity.

\section{Estimation of $\sigma^2_{\Delta x}$ using linear fit}
In cases where the track of the particle can be approximated by a straight line, a linear fit can be made to the x and y projections of the track \cite{pal2012measurement}. The accuracy in the arrival direction  can then be expressed as a function of the fit parameters. The linear fit equation for the x plane projection is:
\begin{equation}
s_x = m_x l + c_x
\label{linearfit}
\end{equation}
$m_x$ and $c_x$ are the slope and intercept of the line, respectively. A similar equation can be written for the y plane projection. Though the roles of strip number and layer number can be swapped in the above equation, this form is chosen for mathematical ease and to avoid problems with infinite slopes in the linear fit. We note from Equation \ref{L} that, to obtain L, the parameters $\Delta x$ and  $\Delta y$ should be known. These in fact can be deduced in terms of the fit parameters using Equation \ref{linearfit} as:
\begin{equation} 
 \Delta s_x = s_{xb}-s_{xt}
 \label{sxerr}
\end{equation}
$s_{xb}$ and $s_{xt}$ are the strip hit positions on the bottom and top layer, respectively, which can be written in terms of the fit parameters as $m_x l_b + c_x$ and  $x_2=m_x l_t + c_x$. $l_t$ is the index of the top layer, which can be written as $N_l-1$ and $l_b$ is the index of the bottom layer (i.e., 0). Therefore, the variance of $\Delta s_x$ is:

\begin{equation}
\sigma^2_{\Delta s_x} = \sigma^2_{m_x} (N_l-1)^2 + 2 \sigma^2_{c_x}
\label{sigdelx}
\end{equation}
Since the linear fit Equation \ref{linearfit} is expressed in terms of dimensionless quantities like the strip number and layer number, Equation \ref{sigdelx} has to be multiplied by a conversion factor $\delta s^2$ to obtain $\sigma^2_{\Delta x}$: 
\begin{equation}
\sigma^2_{\Delta x} = \delta s^2\left\{\sigma^2_{m_x} (N_l-1)^2 + 2 \sigma^2_{c_x}\right\}
\label{sigdelxsq}
\end{equation}

Thus, using Equation \ref{sigdelxsq} in \ref{theory}, we get:
\begin{equation}
 \frac{\sigma^2_{cos(\theta)}}{cos^2(\theta)} = \sigma_f^2\left( \frac {\delta s^2}{(N_l-1)^2 \delta g^2}\right)\sin^2(\theta)cos^2(\theta) 
\label{lerrfit}
\end{equation}
where, 
\begin{equation}
\sigma_f^2 = \sigma^2_{m} (N_l-1)^2 + 2 \sigma^2_{c}
\end{equation}
represents the contribution of the fit parameters to the uncertainty in $cos(\theta)$. Equation \ref{lerrfit} gives the accuracy in the arrival direction in terms of the linear fit parameters, which can be further simplified. For a linear fit, the uncertainties in the fit parameters are given by \cite{bevington2003data}:
\begin{equation}
\sigma_c^2 = \frac{\sigma^2}{\Delta}\Sigma l^2
\label{cerr}
\end{equation}
\begin{equation}
\sigma_m^2 = N_l\frac{\sigma^2}{\Delta}
\label{merr}
\end{equation}
where,
\begin{equation}
\Delta = N_l\Sigma l^2 - (\Sigma l)^2
\label{dellta}
\end{equation}
In the above equations, $\sigma$ denotes the error on the data points. The distribution of the hits on a strip follows a uniform distribution and therefore $\sigma=1/\sqrt{12}$. The summation on $l$ runs from 0 to $N_l-1$ in integer steps and therefore:
\begin{equation}
\Delta =\frac{N_l^2(N_l-1)(2N_l-1)}{6}-\frac{N_l^2(N_l-1)^2}{4}
\end{equation}
which on further simplification gives: 
\begin{equation}
\Delta =\frac{N_l^2(N_l-1)(N_l+1)}{12}
\end{equation}
Substituting the value of $\Delta$ and on simplifying Equations \ref{cerr} and \ref{merr}, we get:
\begin{equation}
\sigma_c^2 =\frac{2(2N_l-1)}{N_l(N_l+1)}(\frac{1}{12})
\end{equation}
and
\begin{equation}
\sigma_m^2 = \frac{12}{N_l(N_l-1)(N_l+1)}(\frac{1}{12})
\end{equation}
and thus,
\begin{equation}
\sigma_f^2=\frac{4(5N-4)}{N_l(N_l+1)}(\frac{1}{12}) 
\label{sigff}
\end{equation}

\section{Validation using odd-even difference method}

Here, we explain the odd-even difference method that was developed based on the existing chessboard method \cite{di2007measurement} to determine the accuracy in the arrival direction estimated using data collected from the IICHEP prototype. In this method, the zenith angle $\theta_{odd}$ is calculated using odd-indexed layers only (l= 1, 3, 5, 7... etc.,) in the fit. Similarly, $\theta_{even}$ is obtained with only the even-indexed layers in the fit. The difference quantity

\begin{equation}
\epsilon = cos(\theta_{even})-cos(\theta_{odd})
\label{diffqty}
\end{equation}
should be centered around zero (i.e., $\langle\epsilon\rangle =0$), if there is no bias between the odd-indexed and even-indexed layers. The variance of $\epsilon$ is given by:
\begin{equation}
\sigma_{\epsilon}^2 = \sigma_{cos(\theta_{even})}^2 + \sigma_{cos(\theta_{odd})}^2
\end{equation}

Since $\sigma_{\epsilon}^2 =  \langle\epsilon^2\rangle -  \langle\epsilon\rangle^2$ and noting that $\langle\epsilon\rangle =0$, we get:
\begin{equation}
 \langle\epsilon^2\rangle = \sigma_{cos(\theta_{even})}^2 + \sigma_{cos(\theta_{odd})}^2
 \label{epsq}
\end{equation}
Also, from Equation \ref{diffqty}, we get
\begin{equation}
\sigma_{\epsilon}^2 = \langle\epsilon^2\rangle = \langle (cos(\theta_{even})-cos(\theta_{odd}))^2\rangle
 \label{epsq2}
\end{equation}

\begin{figure}[H]
	\centering
		\includegraphics[scale=0.30]{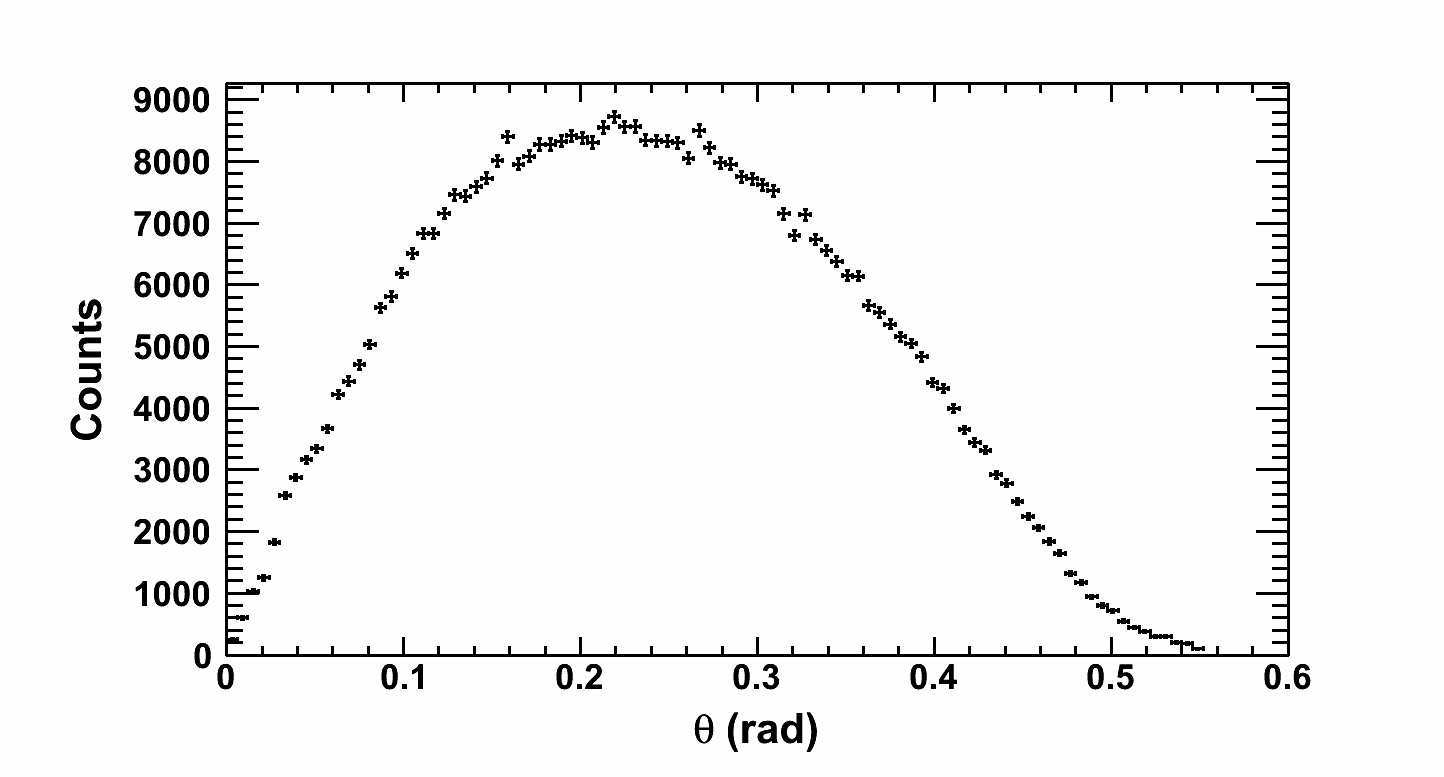}
\caption{The zenith angle distribution of the INO-ICAL prototype detector obtained using a coincidence trigger of the top and bottom layers.}
	\label{zen}
\end{figure}

\begin{figure}[H]
	\centering
		\includegraphics[scale=0.5]{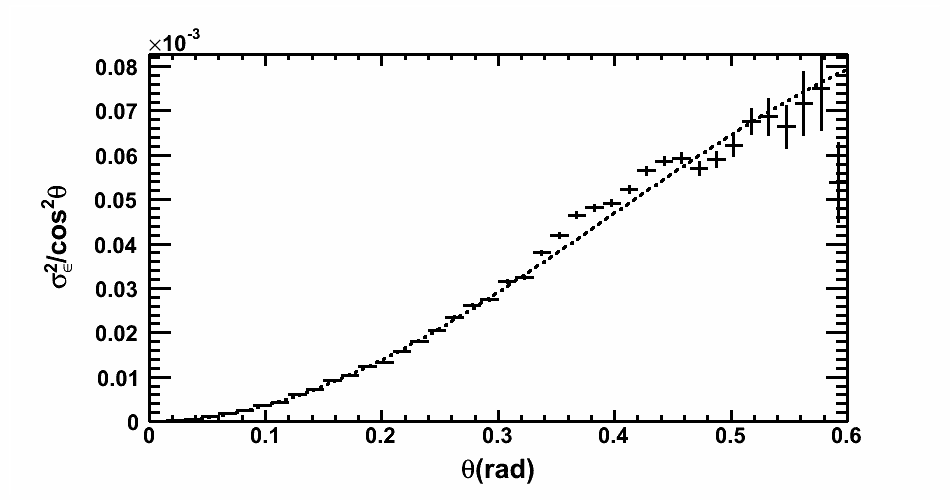}
	\caption{Comparison of accuracy in the arrival direction as computed using Equation \ref{oddeven} (dashed line) and values obtained from data using RHS of Equation \ref{epsq2} (normalized to $cos^2(\theta))$.}
	\label{fig:oddeven}
\end{figure}

Since $\theta_{even}$ and $\theta_{odd}$ are measurable quantities, the RHS of Equation \ref{epsq2} can be estimated from experimental data. Similarly, the RHS of the equation \ref{epsq}
can be calculated from Equation \ref{lerrfit}. Thus, by comparing these results, we validate our theoretical expression.
It is to be noted here that for the analysis using the odd-even difference method, the number of layers for the odd case ($N_{lo}$) and for the even case ($N_{le})$ become half the total number of layers $N_l$, and that the summation in Equations \ref{cerr} and \ref{dellta} will run in even integer steps and odd integer steps, respectively. The layer gap is also to be changed to twice the original value. Thus, Equation \ref{sigff} is to be modified as follows:

\begin{equation}
\sigma_{f(even)}^2=\frac{(11N_{le}-7)}{N_{le}(N_{le}+1)} (\frac{1}{12})
\end{equation}
 
\begin{equation}
\sigma_{f(odd)}^2=\frac{(11N_{lo}^2-6N_{lo}+1)}{N_{lo}(N_{lo}+1)(N_{lo}-1)} (\frac{1}{12})
\end{equation}
Therefore, it follows from Equation \ref{epsq} that
\begin{equation}
\frac{\sigma_{\epsilon}^2}{cos^2(\theta)} = \left\lbrace\sigma_{f(even)}^2 + \sigma_{f(odd)}^2\right\rbrace K
\label{oddeven}
\end{equation}
where,
\begin{equation}
K = \left( \frac {\delta s^2}{(N_l/2-1)^2 (2\delta g)^2}\right)\sin^2(\theta)cos^2(\theta)
\end{equation}
The zenith angle distribution of cosmic muons obtained using this detector with top and bottom layer coincidence trigger is shown in Figure \ref{zen}. \\
The LHS of Equation \ref{epsq2} (normalized to $cos^2(\theta)$) is the same as the LHS of Equation \ref{oddeven}. A comparison between the results obtained using the RHS of Equation \ref{epsq2} (which is the experimental result) and that using Equation \ref{oddeven} is shown in Figure \ref{fig:oddeven}. The figure shows agreement between experimental and theoretical results except for a slight deviation in the range 0.3--0.5 rad, which could not be understood with the existing data. 

\section{Theoretical estimate of angular resolution}
\begin{figure}[H]
	\centering
		\includegraphics[scale=0.9]{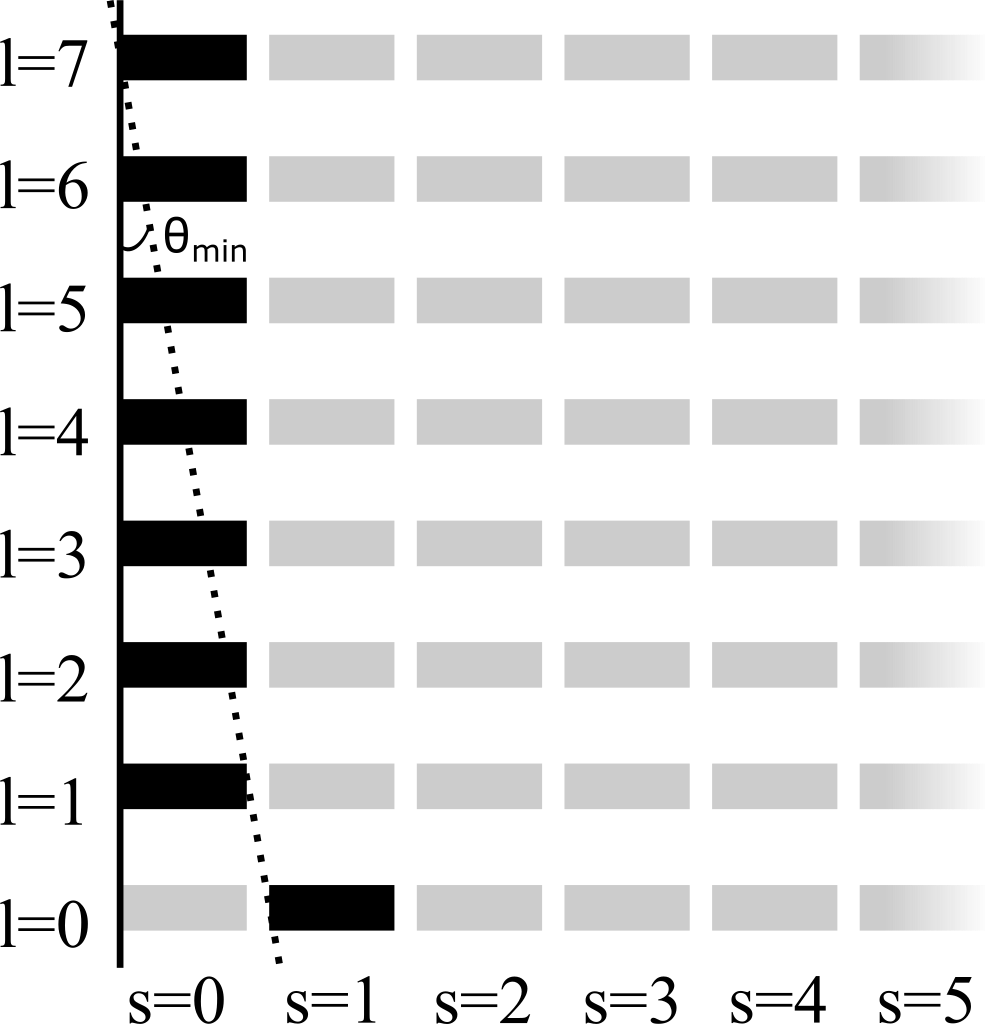}
\caption{The angular resolution can be estimated by considering the least detectable deviation from the straight line as shown in this figure. Only the first 6 strips are shown here for simplicity. The figure is for illustration and not to scale.}
	\label{ang_res}
\end{figure}

Equation \ref{oddeven} in the previous section gives the statistical accuracy of one component (i.e zenith angle) of the arrival direction. The angular resolution is the minimum angular deviation that can be measured using the detector setup. A simple way of estimating this is to consider the least deviation from the straight line detectable with such a detector (Cf. Figure \ref{ang_res}). The figure shows a particle going through the same strip ($s=0$) in all the layers except the bottom layer. In the bottom layer, it hits the adjacent strip ($s=1$). We assume that the track is strictly vertical in the y plane. Any less deviation from this line cannot be measured as all the layers will fire the same strip therefore yielding a zenith angle of zero. Therefore, the angle $\theta_{min}$ this track makes with the vertical can be considered to be the angular resolution of the detector. In making this proposition, we implicitly ignore the effects of multiple hits in a single layer.  From equation \ref{costheta}, the angle for this track can be calculated. For the IICHEP stack, $H$ is 117.6 cm and $L$, the path length, can be estimated using the linear fit parameters for this track. The expressions for fit parameters of a straight line with intercept $c$ and slope $m$ are as follows \cite{bevington2003data}:
\begin{equation}
c = \frac{1}{\Delta}(\Sigma l^2\Sigma s_l - \Sigma l\Sigma ls_l)
\end{equation}
\begin{equation}
m = \frac{1}{\Delta}(N_l\Sigma l s_l - \Sigma l\Sigma s_l)
\end{equation}
\begin{equation}
\Delta = N_l\Sigma l^2 -(\Sigma l)^2
\end{equation}
In the above equations, the conventions laid out in Table \ref{param_table} has been followed. The summations run from $l=0$ to $l=N_l-1$ and $s_l$ is the strip that was hit in layer $l$. The equations can be grossly simplified by taking the case that is considered in Figure \ref{ang_res} (i.e set $s_0=1$ and other $s_l=0$)\footnote{No generality is lost in taking this specific case. Even if all the strip numbers are shifted by 1 (which amounts to shifting the track toward the right), we still end up with the same equation. The calculations are simpler in our special case (in which all but one strip number are zero) since most products in the expressions for slope and intercept vanish.}. The resulting simplified equations for intercept and slope are:
\begin{equation}
c=\frac{2(2N-1)}{N(N-1)}
\end{equation}

\begin{equation}
m=\frac{6}{N(N-1)}
\end{equation}
Using the above equations and Equations \ref{costheta} and \ref{L},
the corresponding angle can be found using the simplified formula:\footnote{In Equation \ref{L} we set $\Delta y$ to be zero as assumed in preceding discussion.}
\begin{equation}
cos(\theta_{min}) = \Big(\Big(\frac{6 \delta s}{N (N-1) \delta g}\Big)^2+1\Big)^{-0.5}
\end{equation}
The estimated angular resolution from the above formula is approximately 15 mrad for the IICHEP stack. As mentioned in \cite{di2007measurement}, the angular resolution can also be estimated by the parameter $\psi_{72}$. $\psi_{72}$ is the limit for which the distribution of difference between $\theta_{even}$ and $\theta_{odd}$ contains 72\% of the events. The angular resolution is then given by $\psi_{72}/1.58$ which for our case yields about 11 mrad. This is not far from our conservative estimate. 

\section{Future scope}
The expression derived in this work, though generic in form, can not be used directly for detectors in which the layer gap is filled with a material. In such cases, factors like multiple coulomb scattering should be considered in the calculations. Systematic effects and the effect of multiple hits have been ignored in this analysis. These should be studied by an analysis of avalanche size and other relevant parameters inside the gas volume. The estimate for the angular resolution is a conservative value which should also be verified in a monte-carlo study. 
\section{Acknowledgments}
The INO project is funded by the Department of Atomic Energy and the Department of Science and Technology, Government of India. The authors would like to thank Mr. Raveendran and Mr. R. R. Shinde at IICHEP, Madurai, for their help with the experiments. %



\begin{thebibliography}{99}

\bibitem{santonico1981development}
Santonico, R., and R. Cardarelli. \emph{"Development of resistive plate counters."} \emph{Nuclear Instruments and Methods in physics research} {\bf 187.2-3} (1981): 377-380.

\bibitem{fonte2002applications}
Fonte, P. \emph{"Applications and new developments in resistive plate chambers."} \emph{IEEE transactions on nuclear science} {\bf 49.3} (2002): 881-887.

\bibitem{mondal2012india}
Mondal, Naba K. \emph{"India-Based Neutrino Observatory (INO)."} \emph{The European 
Physical Journal Plus} {\bf 127.9} (2012): 1-6.

\bibitem{di2007measurement}
Di Sciascio, G., and E. Rossi. \emph{"Measurement of the angular resolution of the ARGO-YBJ detector."} \emph{arXiv preprint} {\bf arXiv:0710.1945} (2007).

\bibitem{ku1966notes}
Ku, H. H. \emph{"Notes on the use of propagation of error formulas."} \emph{Journal of Research of the National Bureau of Standards}  {\bf 70.4} (1966).

\bibitem{pal2012measurement}
Pal, S., et al. \emph{"Measurement of integrated flux of cosmic ray muons at sea level using the INO-ICAL prototype detector."} \emph{Journal of Cosmology and Astroparticle Physics} {\bf  2012.07} (2012): 033.

\bibitem{bevington2003data}
Bevington, Philip R., and D. Keith Robinson. \emph{"Data reduction and error analysis."} \emph{McGraw-Hill} (2003).

\bibitem{majumder2012velocity}
Majumder, G., et al. \emph{"Velocity measurement of cosmic muons using the India-based Neutrino Observatory prototype detector."} \emph{Nuclear Instruments and Methods in Physics Research Section A: Accelerators, Spectrometers, Detectors and Associated Equipment} {\bf 661} (2012): S77-S81.
\bibitem{bhuyan2012vme}
 Bhuyan, M., et al. \emph{"VME-based data acquisition system for the India-based Neutrino Observatory prototype detector."} \emph{Nuclear Instruments and Methods in Physics Research Section A: Accelerators, Spectrometers, Detectors and Associated Equipment} {\bf 661} (2012): S73-S76.
\bibitem{behere2013electronics}
Behere, Anita, et al. \emph{"Electronics and data acquisition system for the ICAL prototype detector of India-based neutrino observatory."} \emph{Nuclear Instruments and Methods in Physics Research Section A: Accelerators, Spectrometers, Detectors and Associated Equipment} {\bf 701} (2013): 153-163.

\bibitem{satya_thesis}
Satyanarayana, Bheesette. \emph{"Design and characterisation studies of resistive plate Chambers."} \emph{Diss. PhD thesis}, Department of Physics, IIT Bombay, {\bf PHY-PHD-10-701}, 2009.






\end{thebibliography}
\end{document}